\documentclass[a4paper,11pt]{article}
\usepackage[utf8x]{inputenc}
\usepackage{graphicx}
\usepackage{url}
\usepackage{comment}
\usepackage[portuguese,algoruled,longend]{algorithm2e}
\usepackage[top=1.6cm, bottom=1.6cm, left=1.6cm, right=1.6cm]{geometry}
\usepackage{textcomp}
\usepackage{url}
\usepackage{natbib}
\usepackage{color}
\usepackage{subfigure}

\newcommand\blfootnote[1]{%
  \begingroup
  \renewcommand\thefootnote{}\footnote{#1}%
  \addtocounter{footnote}{-1}%
  \endgroup
}

\begin{document}

\begin{center} \begin{Large} \textbf{Can WhatsApp Benefit from Debunked Fact-Checked Stories to Reduce Misinformation?*\blfootnote{*This is a preprint version of an accepted manuscript on The Harvard Kennedy School (HKS) Misinformation Review. Please, consider to cite it instead of this one.}}
\end{Large} \end{center} 

\begin{normalsize}
	\begin{center}
\normalsize\textbf{{Julio C. S. Reis$^{\S}$, Philipe Melo$^{\S}$, Kiran Garimella$^{\ddag}$, Fabr\'icio Benevenuto$^{\S}$}} \\
	$^{\S}$Computer Science Department, Universidade Federal de Minas Gerais (UFMG), Brazil\\
   $^{\ddag}$Massachusetts Institute of Technology (MIT), United States\\
    {\{julio.reis, philipe, fabricio\}@dcc.ufmg.br}, 
	{\{garimell\}@mit.edu}\\ 
\vspace{0.2cm}

	\end{center}
\end{normalsize}

\section*{Abstract} \vspace*{-0.2cm}

\noindent\textit{WhatsApp was alleged to be widely used to spread misinformation and propaganda during elections in Brazil and India. Due to the private encrypted nature of the messages on WhatsApp, it is hard to track the dissemination of misinformation at scale. In this work, using public WhatsApp data, we observe that misinformation has been largely shared on WhatsApp public groups even after they were already fact-checked by popular fact-checking agencies. This represents a significant portion of misinformation spread in both Brazil and India in the groups analyzed. We posit that such misinformation content could be prevented if WhatsApp had a means to flag already fact-checked content. To this end, we propose an architecture that could be implemented by WhatsApp to counter such misinformation. Our proposal respects the current end-to-end encryption architecture on WhatsApp, thus protecting users’ privacy while providing an approach to detect the misinformation that benefits from fact-checking efforts.}

\section*{Research Questions} 

\begin{itemize}
    \item To what extent is misinformation shared on WhatsApp after they are debunked? 
    \item How can WhatsApp benefit from using fact-checked content in the fight against misinformation?
\end{itemize}

\section*{Essay Summary} 

\begin{itemize}
    \item We collected two large public datasets consisting of: (i) images from public WhatsApp groups during recent elections in Brazil and India; and (ii) images labeled as misinformation by professional fact-checking agencies in these two countries. This data was used to perform a series of experiments to evaluate if debunked fact-checked stories are shared on WhatsApp;
    \item By looking at the time when an image was shared on WhatsApp and when it was debunked by fact-checkers, we find that a non-negligible amount of the fact-checked images in our WhatsApp dataset.  Particularly, we show that 40.7\% and 82.2\% of the shares containing misinformation in our sampled data from Brazil and India were made after the content was debunked by the fact-checking agencies.  This suggests that, if WhatsApp flagged the content as false at the time it was fact-checked, it could prevent a large fraction of shares of misinformation from happening. Given that rumors and falsehoods have led to lynchings and social unrest in many countries in the global south~\citep{banaji2019whatsapp}, such solutions are the need of the hour;
    \item Building on this idea, we discuss a potential architecture change on WhatsApp that would allow it to benefit from fact-checked content to counter misinformation. Our proposal is based on on-device fact-checking, where WhatsApp can detect when a user shares content that has been previously labeled as misinformation by fact-checkers by using a digital fingerprint for that content. The potential architecture makes it possible for WhatsApp to protect users' privacy, keep the end-to-end encryption, and detect misinformation in a timely fashion.
\end{itemize}

\section*{Argument \& Implications} 

Social media platforms have dramatically changed how people consume and share news. An individual user can reach as many readers as other traditional media nowadays \citep{allcott2017social}. Social communication around news is becoming more private as messaging apps continue to grow around the world. With over 2 billion users, WhatsApp is a primary network for discussing and sharing news in countries like Brazil and India where smartphone use for news access is already much higher than other devices, including desktop computers and tablets~\citep{newman2019reuters}. WhatsApp users send more than 55 billion messages a day, of which 4.5 billion are images~\citep{blog2017connecting}. Along with massive information flows, a large amount of misinformation is posted on WhatsApp without any moderation~\citep{resendewww19}. Several works have already shown how misinformation has negatively affected the democratic discussion in some countries~\citep{moreno2017whatsapp, resendewww19} and even led to violent lynchings~\citep{arun2019whatsapp}. Furthermore, during the COVID-19 pandemic, there has been an uptick in rumours and conspiracies spreading through online platforms\citep{ferrara2020types}. The International Fact-Checking Network found more than 3,500 false claims related to COVID-19 in less than two months \citep{poynter2020}. WhatsApp has played a big role in this infodemic around the world \citep{delcker2020, romm2020}. Several high profile examples, including state-backed misinformation campaigns on WhatsApp have been observed recently \citep{brito2020}. Unlike social platforms such as Twitter and Facebook, which can perform content moderation and hence take down offending content, the end-to-end encrypted (E2EE) structure of WhatsApp creates a very different scenario where the same approach is not possible. Only the users involved in the conversation have access to the content shared, shielding abusive content from being detected or removed. This obstacle has already led governments to propose breaking encryption as a way to enable moderation and law enforcement on platforms like WhatsApp~\citep{sabbagh2019}. The key challenge we tackle in this paper is to find a balance between fighting misinformation on WhatsApp and keeping it secure with end-to-end encryption.

We first assess the role of data from professional fact-checking agencies to combat misinformation on WhatsApp. We measure this using data collected from both fact-checking agencies and public groups on WhatsApp during recent elections in Brazil and India. In our analysis, we found that a significant part of misinformation shared in the groups we monitor were done after the content was fact-checked as misinformation by the fact-checking agencies. This represents a significant fraction of shares of misinformation, 40.7\% and 82.2\% in Brazil and India respectively. This shows that, even though fact-checking efforts label content as misinformation, it is still being freely shared on the platform, indicating that public fact-checking alone does not block the spreading of misinformation on WhatsApp. This is a missed opportunity for WhatsApp, which could have prevented the further spread of already known misinformation.

Building upon this finding, we propose a moderation methodology that benefits from the observed potential of using fact-checking, in which WhatsApp can automatically detect when a user shares images and videos which have previously been labeled as misinformation using hashes, without violating the privacy of the user nor compromising the E2EE within the messaging service. This solution is similar to how Facebook would flag content as misinformation, based on having hashes of previously fact-checked content. The proposal works entirely on the device of the user before the content is encrypted, thus making it tenable within the existing E2EE design on WhatsApp. 

\begin{figure}[ht!] 
	\centering
	   \includegraphics[width=0.8\linewidth]{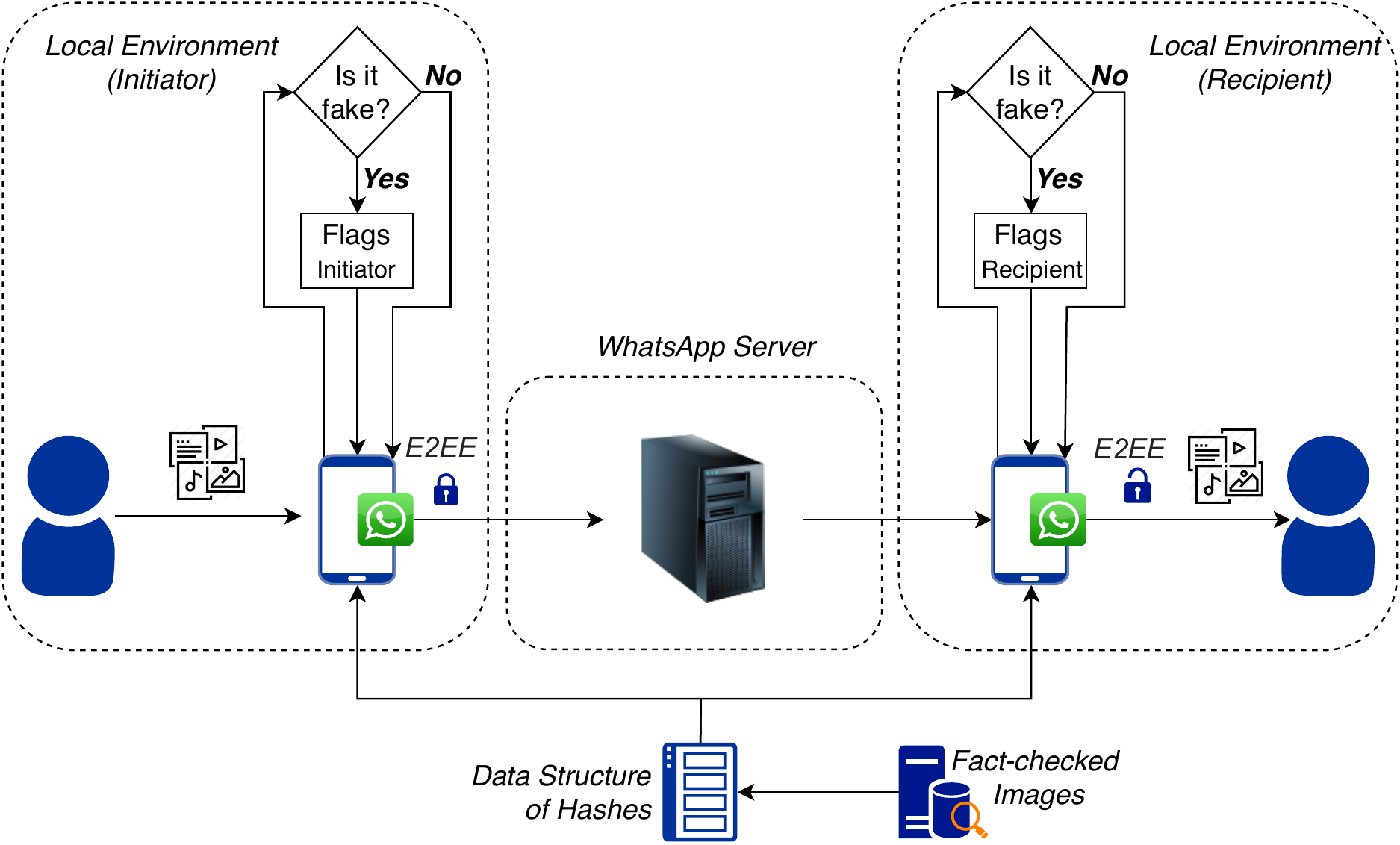}
	\caption{Proposed architecture for on-device fact-checking.}
	\label{fig:proposed_architecture_2}
\end{figure}

An overview of the potential architecture is shown in Figure~\ref{fig:proposed_architecture_2} and can be explained in the following steps: (i) WhatsApp maintains a set of hashes of images which have been previously fact-checked, either from publicly available sources or through internal review processes. (ii) These hashes are shipped with the WhatsApp app, storing it on a user’s phone. This step can be periodically updated based on images that Facebook’s moderators have been fact-checking on Facebook, which is much more openly accessible. This set could be condensed and efficiently stored using existing probabilistic data structures like Bloom Filters \citep{song2005fast}. (iii) Once a user intends to send an image, WhatsApp checks whether it already exists in the hashed set on the user’s device. If so, a warning label could be applied to the content or further sharing could be denied. (iv) The message is encrypted and transferred through the usual E2EE methods. (v) When the recipient user receives the message, WhatsApp decrypts the image on the phone, obtains a perceptual hash, and also checks it on a hashed set on the receiver’s end. (vi) If it already exists, the content is flagged on the receiver's end. The flagging could be implemented as a warning shown to the user indicating that the image could be potential misinformation, providing information about where the image was fact-checked; and in addition, also prevent the image from being forwarded further.

The strategy works to prevent coordinated disinformation campaigns that are particularly important during elections and other high profile national and international events (such as the COVID-19 crisis) but also stops basic misinformation, where a lack of awareness leads to spreading. For instance, we observed that roughly 15\% of the images in our data were related to false health information. These are forwarded mostly with the assumption that they might help someone in case they are true. In some cases (e.g. the child kidnapping rumors \citep{arun2019whatsapp}, such benign forwarding of misinformation leads to violence and killings.

The potential architecture requires changes in WhatsApp, as it introduces a new component containing hashes stored on the phone and also checking images. It provides high flexibility and the ability to detect near similar images, hence increasing the coverage and effectiveness in countering misinformation. As previously mentioned, this architecture also fully abides by the current E2EE pipeline WhatsApp has, where WhatsApp does not have access to any content information. All the matching and intervention is done on the device without the need for any aggregate metadata in the message. Facebook, that owns WhatsApp, could optionally keep statistics of how many times a match occurred to establish the prevalence and virality of different types of misinformation and to collect stats about users who repeatedly send such content. Note that similar designs have been proposed recently in informing policy decisions in light of governments requesting a backdoor in the encryption~\citep{gupta2018, mayer2019moderation}.

It is important to mention that while WhatsApp messages are encrypted in transit, the endpoint devices, such as smartphones and 
computers, do not offer encryption. In this sense, this potential architecture adds new components to the client, adding also more potential for security breaches. Our idea is practical and deployable because it is an industry-standard to detect unlawful behavior in social media platforms~\citep{Farid2018abuses}. For example, WhatsApp scans all unencrypted information on its network such as user/group profile photos, and group metadata for unlawful content such as child pornography, drugs, etc. If flagged, these are manually verified~\citep{constine2018} and the abusing accounts are banned. Our proposal extends the same methodology to the user’s device in order to enable private detection and it could be easily integrated within a network of fact-checkers that Facebook already has up and running.

Guaranteeing the privacy of the users is a goal as essential as combating misinformation. In our understanding, both can coexist in parallel in the EE2E encrypted chat environment of WhatsApp and our objective is to provide a middle ground solution in between that could satisfy those who request actions in combat of misinformation spreading in such platforms, but also keep the privacy of on the users' phone before it is encrypted.

Most previous qualitative research on misinformation in the form of images on social media~\citep{paris2019cheap,brennen2020, garimella2020images} has found that “cheap fakes” or old images reshared out of context are the most popular form of misinformation. Our work proposes the use of a database to maintain hashes (fingerprints) of such old images which were already fact-checked and use the database to prevent the spread of misinformation without breaking encryption on WhatsApp. Our work shows that a significant amount of shares of highly popular misinformation images could have been prevented if our proposal was applied during the Indian and Brazilian elections. The current debate on content moderation on WhatsApp has been only at either of two extremes: On the one hand, governments claiming that WhatsApp provides them with a backdoor, essentially breaking encryption, and on the other hand, WhatsApp claiming that any such moderation is impossible because of the end to end encryption design. This work provides a concrete suggestion to WhatsApp on what could be done. Since misinformation on WhatsApp has led to tragic real-life consequences like lynchings, social disorder, and threats to democratic processes, implementing such a system, even on a case by case basis (such as during elections or a pandemic) must be considered. This work also adds a counter-voice to claims of breaking encryption on WhatsApp by showing that a simple solution might go a long way in addressing the misinformation problem on WhatsApp. Since a majority of users never share viral misinformation, this design change, even when applied to all WhatsApp users, would be perceived by, and potentially affects a small set of users on WhatsApp, although it could still have a high impact in detecting and preventing the spread of misinformation.

It is important to mention that there might be multiple design choices to deal with a content that is detected as misinformation by WhatsApp~\citep{nyhan2010corrections, clayton2019real}, and user experience tests need to be conducted to investigate the best of them. For instance, WhatsApp could inform the user that the content was debunked, but still allow it to spread, or prevent it to be forwarded, block users, etc. This is a decision that WhatsApp should make, consulting previous research on the effects of flagging content as misinformation ~\citep{nyhan2010corrections, clayton2019real}. 

\section*{Findings} 

\noindent \textit{We identified images which were fact-checked to be false by popular fact-checking agencies in India and Brazil and analyzed the sharing patterns of these images on public WhatsApp groups. We found that  40.7\% and 82.2\% of the shares of these misinformation images from Brazil and India respectively, occurred after the images were debunked by fact-checking agencies.  This indicates that the fact-checking efforts do not completely stop the spread of misinformation on WhatsApp. This also provides an opportunity for a large fraction of misinformation that could be stopped by preventing those shares using previously fact-checked data.}
\\

Our analysis requires WhatsApp messages containing misinformation including their propagation, as well a set of content labeled as misinformation by fact-checkers in order to assess the potential of fact-checking to combat misinformation. Thus, we collected data from over 400 and 4,200 public groups from Brazil and India, respectively, dedicated to political discussions. We also performed an extensive collection of hundreds of images that were fact-checked by popular fact-checking agencies in India and Brazil. Details of this collection are provided in the Methods section.

After matching the occurrences of misinformation labeled by professionals with those shared in  WhatsApp, we evaluated the volume of misinformation shared in our sample after the time it was already debunked. To get estimates on these shares, we measured the occurrence of these fact-checked images in our WhatsApp. This way, we are able to measure how many posts were done for each misinformation image before and after the first fact-check of this image.

\begin{figure}[t]
	\centering
	\subfigure[Brazil]{\includegraphics[width=0.37\linewidth]{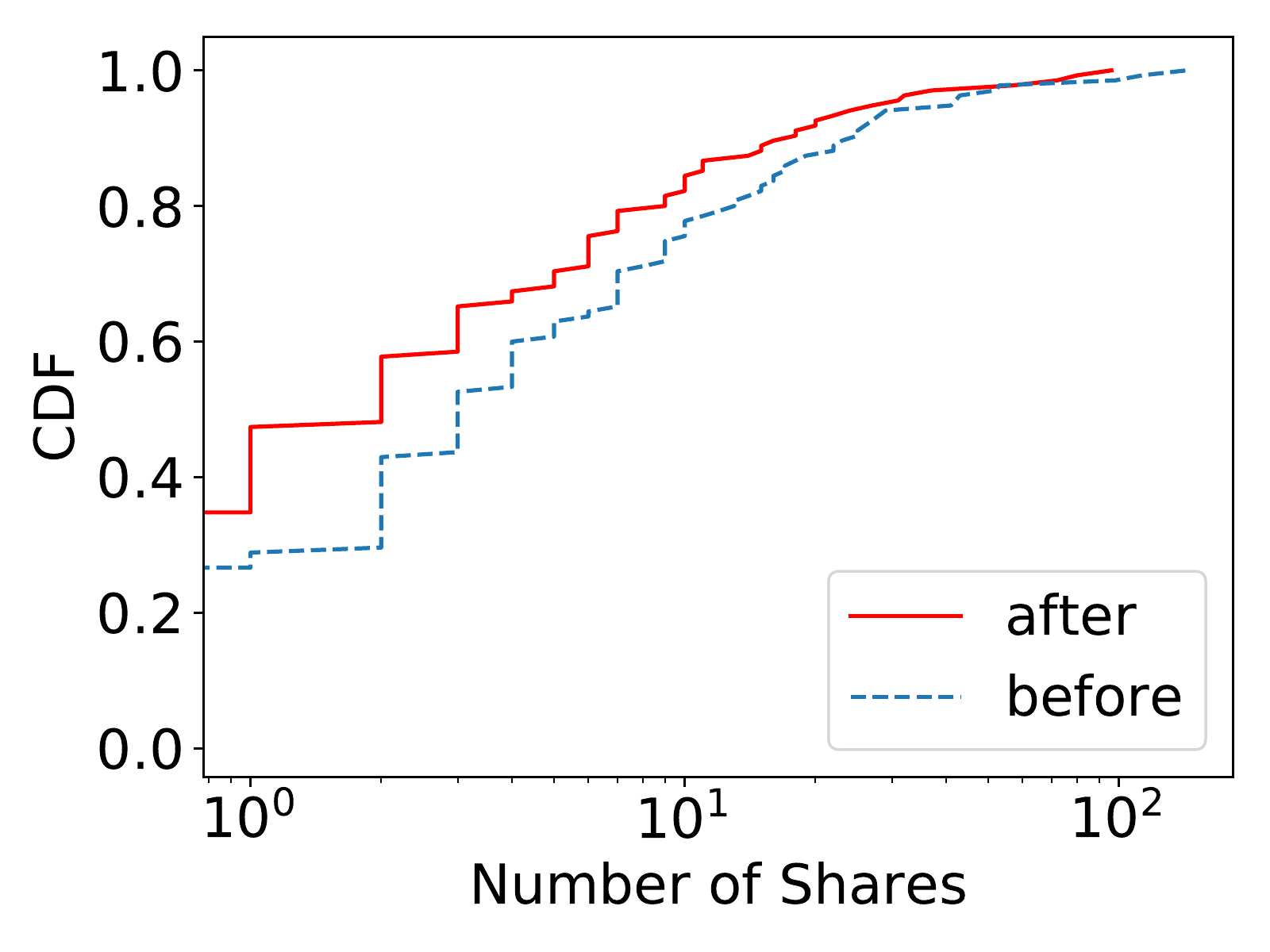}\label{fig:shares_BRAZIL}}
	\subfigure[India]{\includegraphics[width=0.37\linewidth]{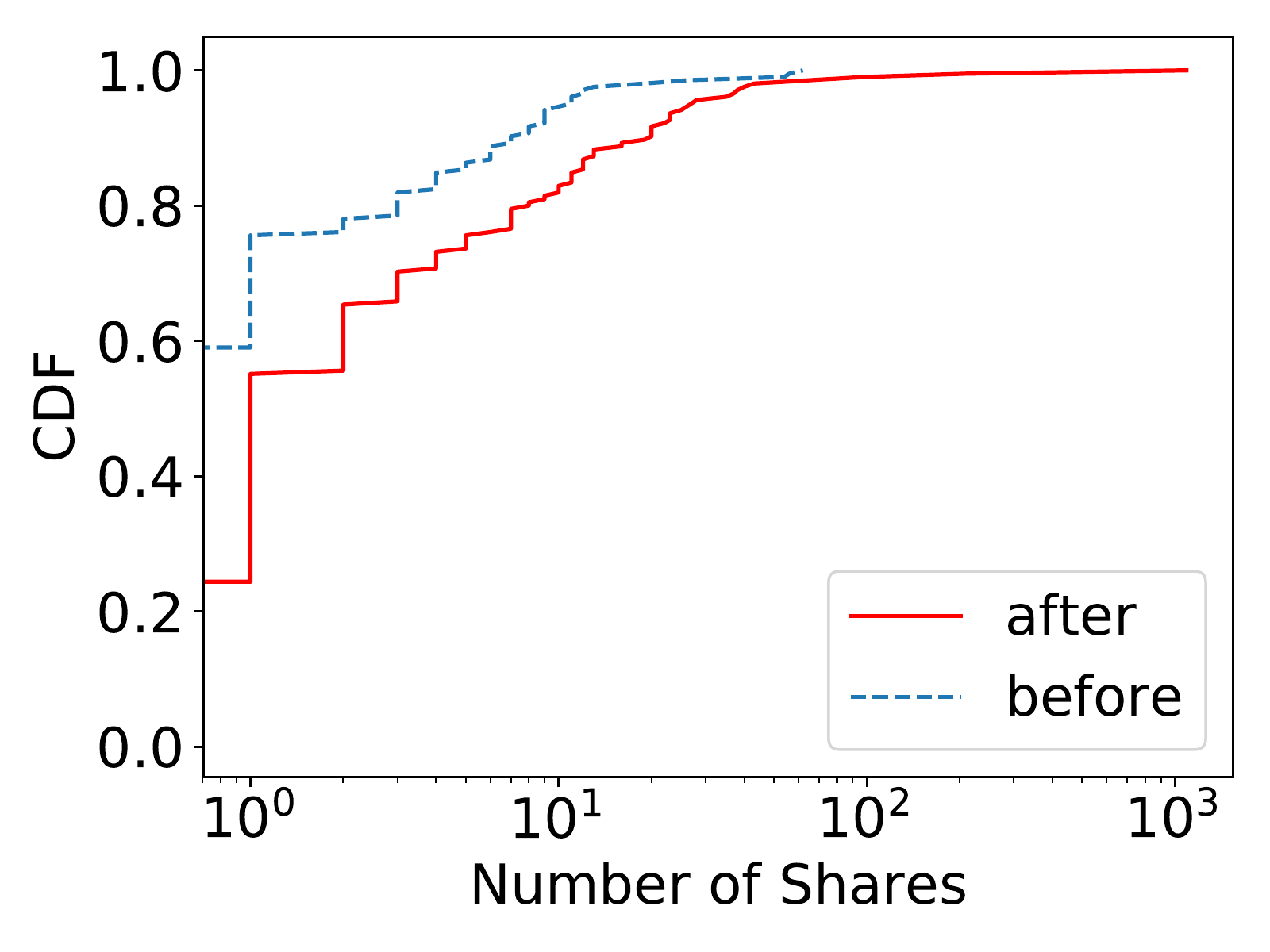}
	\label{fig:shares_INDIA}}
	\caption{Cumulative distributions of shares of images containing misinformation before/after fact-checking on both countries.} 
	 \label{fig:before_after_shares}
	 	\vspace{-0.3cm}
\end{figure}

Figure~\ref{fig:before_after_shares} shows the cumulative distribution function (CDF) of the number of shares done before and after the checking for each distinct misinformation image. We can observe in both countries that for the most broadly shared images there are as many posts before as after the checking date. Moreover, for India, there are more shares after checking than before and there are even images with up to 1,000 shares after fact-checking while the maximum shares before do not exceed 100. 

As shown in Table~\ref{tab:potential}, summing all shares on our sampled data, we find that 40.7\% of the misinformation image shares in Brazil and 82.2\%\footnote{This number drops to 71.7\% if we remove the outlier image with the maximum number of shares, as it was shared over 1000 times.} of the shares in India happened after we already know the content is false. We observe that a considerable number of images are still shared on WhatsApp, even though they were already publicly fact-checked. This demonstrates that fact-checking content on WhatsApp is, to a large extent, a one-sided effort and is not enough to combat the huge volume of misinformation shared on an encrypted, closed platform like WhatsApp.

\begin{table}[]
\centering
\caption{Number of images containing misinformation shared in our WhatsApp dataset from Brazil and India and comparison of shares before and after the checking date of fact-checking agencies.}
\label{tab:potential}
\begin{tabular}{l|l|l|l|l|}
\cline{2-5}
                             & \multicolumn{1}{c|}{\begin{tabular}[c]{@{}c@{}}Misinformation\\ Images found\end{tabular}}  & \multicolumn{1}{c|}{\begin{tabular}[c]{@{}c@{}}Total \\ Shares\end{tabular}} & \multicolumn{1}{c|}{\begin{tabular}[c]{@{}c@{}}\%Shares\\ After\\ Checking\end{tabular}} & \multicolumn{1}{c|}{\begin{tabular}[c]{@{}c@{}}Max Shares\\ After\\ Checking\end{tabular}} \\ \hline
\multicolumn{1}{|l|}{Brazil} & 135                                                                                                                                                                         & 2,209                                                                        & 40.7\%                                                                                     & 96                                                                                         \\ \hline
\multicolumn{1}{|l|}{India}  & 205                                                                                                                                                                          & 2,944                                                                        & 82.2\%                                                                                     & 1,089                                                                                      \\ \hline
\end{tabular}
\vspace{-\baselineskip}
\end{table}

\section*{Methods} 

Combating misinformation in messaging apps such as WhatsApp is a big challenge due to end to end encryption (E2EE). This is even more challenging in an infodemic where we’ve been experiencing an avalanche of misinformation spreading on WhatsApp~\citep{delcker2020}. The main goal of this work is to investigate the impact of fact-checking in detecting misinformation in such systems and to propose an architecture capable of automatically finding and counter misinformation shared on WhatsApp and systems protected with E2EE safety without violating users’ privacy.

For this study, we need a large dataset from WhatsApp containing misinformation and a large dataset from fact-checkers identifying which content is fake. To obtain WhatsApp data, we used existing tools~\citep{garimella2018whatapp} to collect data from public WhatsApp groups\footnote{Any WhatsApp group with a link to join publicly available is considered a public group.}. We specifically were interested in monitoring WhatsApp conversations discussing politics in two of WhatsApp’s biggest markets: India and Brazil. We first obtained a list of publicly accessible WhatsApp groups by searching extensively for political parties and politicians in the two countries on Facebook and Google. We obtained over 400 and 4,200 groups from Brazil and India respectively, dedicated to political discussions. Note that even though we were thorough in our search, these groups at best represent a convenience sample of all WhatsApp groups. The period of data collection for both countries includes the respective national elections in these countries\footnote{Elections in India and Brazil took place over a few weeks. In India, they were held from 11 April to 19 May 2019, and from 7 to 28 October 2018 in Brazil. We started our data collection 2 months before the start of the elections and stopped the collection 1 month after the election.}. Public groups discussing politics have been shown to be widely used in both these countries~\citep{newman2019reuters,lokniti2018} and contain a large amount of misinformation~\citep{resendewww19,melo2019whatappMonitor}. For this work, we choose to filter only messages containing images, which make up over 30\% of the content in our dataset. The dataset overview and the total number of users, groups, and distinct images are described in Table~\ref{tab:dataset}. Note that the volume of content for India is ten times bigger than Brazil.

\begin{table}[ht!]
\caption{Statistics of our WhatsApp data.}
\centering
\label{tab:dataset}
\begin{tabular}{l|l|l|l|l|l|}
\cline{2-6}
                             & \multicolumn{1}{c|}{\#Users} & \multicolumn{1}{c|}{\#Groups} & \multicolumn{1}{c|}{\begin{tabular}[c]{@{}c@{}}Unique\\ Images\end{tabular}} & \multicolumn{1}{c|}{\begin{tabular}[c]{@{}c@{}}Total\\ images\end{tabular}} & \multicolumn{1}{c|}{Time Span} \\ \hline
\multicolumn{1}{|l|}{Brazil} & 17,465                       & 414                           & 4,524                                                                        & 34,109                                                                          & 2018/08 - 2018/11     \\ \hline
\multicolumn{1}{|l|}{India}  & 63,500                          & 4,250                            & 509k                                                                         & 810k                                                                      & 2019/02 - 2019/06     \\ \hline
\end{tabular}
\vspace{-\baselineskip}
\end{table}

Since most of WhatsApp is private and obtaining data is hard, public groups have become a common approach to evaluate information flows on WhatsApp \citep{garimella2018whatapp, resendewww19,melo2019whatappMonitor}.  Although this represents a small portion of messages exchanged on all of WhatsApp, collecting WhatsApp data at scale is a challenge itself due to its closed nature. To the best of our knowledge, the datasets used in this study are the largest in understanding the spread of misinformation images on WhatsApp.

To collect a set of misinformation pieces that already spread we obtained images that were fact-checked in the past by fact-checking agencies for each country. First, we crawl all images which were fact-checked from popular fact-checking websites from Brazil and India. For each of these images, we also obtained the date when they were fact-checked. Second, we used Google reverse image search to check whether one of the main fact-checking domains were returned when searching for an image in our database. If so, we parsed the fact-checking page and automatically labeled the image as fake or true depending on how the image was tagged on the fact-checking page~\citep{resendewww19}. Our final dataset of images previously fact-checked contains 135 images from Brazil and 205 images from India, which were shown to contain misinformation. It is important to highlight that many checking agencies do not post the actual image that has been disseminated. Often only altered versions of the image are posted and other versions of the false story are omitted to avoid contributing to the spreading of misinformation. This leads to us to have a small number of matches compared to the total number of fact-checked images we obtained, but that is sufficient to properly investigate the feasibility of the potential architecture. Note that even though the set of fact-checked images was small, the fact that these images have been fact-checked means that they were popular and spread widely. Table 1 shows a summary of the fact-checked images and their activity in our dataset. 

Next, we used a state-of-the-art perceptual hashing based image matching technique, PDQ hashing\footnote{https://github.com/facebook/ ThreatExchange/tree/master/hashing/pdq/java}, to look for occurrences of the fact-checked images in our WhatsApp data. The PDQ hashing algorithm is an improvement over the commonly used pHash~\citep{zauner2010implementation} and produces a 256-bit hash using a discrete cosine transformation algorithm. PDQ is used by Facebook to detect similar content and is the best known state-of-the-art approach for clustering together similar images. The hashing algorithm can detect near similar images, even if they were cropped differently or they have small amounts of text overlaid on them. Finally, not all images which are fact-checked contain misinformation. To make sure our dataset was accurately built, we manually verified each image that appears in both the fact-checking websites and in the WhatsApp data.

Our WhatsApp data originated only for public WhatsApp groups, and we emphasize that all sensitive information (i.e., group names and phone numbers) were anonymized in order to ensure the privacy of users. The dataset only consists of images that were publicly fact-checked and anonymized user/group information. Hence, to the best of our knowledge, it does not violate the WhatsApp terms of service.

Our methodology has the following limitations: (i) Labeling and implementing forwarding restrictions on already known fake images can only help to a certain degree. There is inconclusive evidence on whether labeling images as misinformation might backfire~\citep{nyhan2010corrections, levin2017,wood2019elusive}; (ii) Our dataset of WhatsApp groups is not representative. However, since most of WhatsApp data is private, any research or evaluation of a policy proposal must be done on the publicly available data from WhatsApp. So, even though our data is limited, it provides hints at sharing patterns of misinformation on WhatsApp. To date, this is the largest WhatsApp data available sample; (iii) On a similar note, the images from fact-checking websites are also a biased sample. Notably, the fact that these images have been selected by a fact-checking agency to be debunked means that they were already popular on social media. (iv) Finally, our proposed solution is only one piece in solving the misinformation problem. It should not be considered as a one-step solution to the problem of misinformation as it also relies on fact-checking efforts done by third parties experts. 

\newpage

\bibliographystyle{apalike}
\vspace*{-0.2cm}
\bibliography{references}

\newpage

\section*{Funding} 
This research was partially supported by Ministério Público de Minas Gerais (MPMG), project Analytical Capabilities, as well as grants from FAPEMIG, CNPq, and CAPES. Kiran Garimella was funded by a Michael Hammer postdoctoral fellowship at MIT.
\section*{Competing interests} 
The author(s) declared no potential conflicts of interest with respect to the research, authorship, and/or publication of this article.
\section*{Ethics} 
This work does not involve any human subjects. The fact-checking data is publicly available and only includes highly shared images. WhatsApp data collection for India was approved by MIT's Committee on the Use of Humans as Experimental Subjects, which included a waiver of informed consent.
\section*{Copyright} 
This  is  an open access article distributed under the terms of the \underline{Creative  Commons  Attribution  License}, which permits unrestricted use, distribution, and reproduction in any medium, provided that the original author and source are properly credited.
\section*{Data availability} 
All the data used in this paper is publicly available here: \url{ http://doi.org/10.5281/zenodo.3779157} \citep{reis2020dataset}.

\end{document}